\documentclass[11pt,a4paper]{article}
\usepackage[T1]{fontenc}
\usepackage[utf8]{inputenc}
\usepackage{lmodern,microtype}
\usepackage[margin=25mm]{geometry}
\usepackage{graphicx,booktabs,tabularx,array,amsmath}
\usepackage[section]{placeins}
\graphicspath{{figures/}}
\usepackage[dvipsnames]{xcolor}
\usepackage[colorlinks=true,linkcolor=MidnightBlue,citecolor=MidnightBlue,urlcolor=MidnightBlue]{hyperref}
\usepackage[font=small,labelfont=bf]{caption}
\usepackage{enumitem}
\usepackage{fancyhdr}
\setlist{nosep,leftmargin=*}

\newcolumntype{Y}{>{\raggedright\arraybackslash}X}
\title{\textbf{Critsly and StudioCrit:}\newline An Artefact-Aware AI Critique Workspace\newline and Simulation-Based Readiness Study\newline for Design Education}
\author{Nizam Kadir\\\small Science, Mathematics and Technology\\\small Singapore University of Technology and Design, Singapore\\\small\href{mailto:nizam_kadir@mymail.sutd.edu.sg}{\texttt{nizam\_kadir@mymail.sutd.edu.sg}}}
\date{}
\hypersetup{pdftitle={Critsly and StudioCrit: An Artefact-Aware AI Critique Workspace and Simulation-Based Readiness Study for Design Education},pdfauthor={Nizam Kadir}}
\begin{document}
\maketitle
\begin{abstract}
Critique in design education depends on interpreting work in progress, articulating intentions and translating feedback into revisions. This technical report presents Critsly, an artefact-aware AI critique workspace, and StudioCrit, its architecture-studio research mode. Critsly combines a visual board, design-intention fields, guided reflection, perspective-based critique and action planning. StudioCrit adds studio/class organisation, role-based access, cognitive and architectural classification, educator analytics and exportable evidence. The report consolidates implementation and simulation evidence recorded in a research project submitted in July 2026. Three simulated studio scenarios yielded 109 classified evidence rows, including 85 assigned to higher-order Bloom categories. A separate rehearsal using 50 disposable learner accounts yielded 56 evidence rows, including 46 assigned to higher-order categories. A subsequent hardening rehearsal recorded 50 completed sessions, 50 successful board pulls and 50 denials of student access to analytics. These are software and synthetic-trace observations, not measurements of learning gains or human cognitive performance. Automated classifications remain provisional, and the source report does not establish classifier accuracy or inter-rater reliability. The contribution is an implemented critique-to-evidence workflow and a bounded account of its readiness for further controlled evaluation.
\end{abstract}
\textbf{Keywords:} educational AI; design critique; architecture studio; artefact-aware systems; learning analytics; simulation; research software.

\section{Introduction}
Design critique connects an evolving artefact with the learner's intentions and the judgement of others. Studio conversations convey disciplinary knowledge through discussion of work in progress \cite{uluoglu,goldschmidt}. A useful digital critique environment must therefore do more than retain a sequence of prompts and responses: it should preserve the object of discussion, make relevant context accessible and connect proposed changes to subsequent work.

Critsly addresses this requirement through a shared visual workspace. Boards contain design intentions, notes, media, annotations and links; AI critique uses structured board context together with the learner's request. Guided reflection and selectable critique perspectives support the transition from discussing an artefact to identifying possible revisions. StudioCrit extends this workspace with class organisation and evidence capture for architecture studio research. It attaches provisional cognitive and architectural labels to critique notes and makes their distributions available for inspection.

This paper consolidates implementation and simulation evidence documented during the author's research project in July 2026 \cite{report}. Observations describe the implementation and rehearsals documented at that time. Reported counts and timings are transcribed from the project report; no new classroom experiment or reconstruction of the underlying event logs is claimed.

The research contribution comprises (1) integration of artefact context, structured critique and action planning; (2) a studio/class and evidence workflow supporting distinct learner, educator and researcher roles; and (3) simulation and deployment-readiness observations that identify what remains unvalidated. The emphasis is engineering and AI orchestration within the collaborative Critsly/StudioCrit project, not independent authorship of its broader studio pedagogy.

\subsection{Relationship to the existing Critsly demo paper}
The AIED 2026 Interactive Events contribution by Kadir, Salazar Rodriguez and Ali presents the Critsly demonstration workflow and is available as arXiv:2607.09673 \cite{critsly}. The present report documents the author's coursework engineering contribution and expands the account to include StudioCrit, simulated critique distributions, research export fields, role checks and deployment hardening. It neither replaces that coauthored demo paper nor reports a newly completed human-subject evaluation. Conference demo acceptance should not be interpreted as validation of the classifier or evidence of learning effectiveness.

\section{Research Aim and Design Foundations}
The project aimed to implement and examine an AI-supported critique environment in which learners work around design artefacts and educators can inspect the resulting evidence. Four questions organised the engineering work:
\begin{enumerate}[label=RQ\arabic*.]
\item Can the critique workflow retain board context, design intentions, notes, media and critique history?
\item Can educators organise studios/classes with invitations, multiple linked boards and role-based access?
\item Can the system expose cognitive and architectural classifications in a form that educators and researchers can inspect?
\item What evidence of workflow operation, export completeness and staging readiness is available before a live research pilot?
\end{enumerate}
These questions concern implemented capabilities and observability. They do not test whether AI feedback improves design quality, whether labels identify actual cognition, or whether classroom outcomes exceed those of another teaching approach.

Studio-critique research motivates keeping interpretation connected to student work \cite{uluoglu,goldschmidt}; reflective practice similarly foregrounds the relationship between action, interpretation and revision \cite{schon}. Bloom's revised taxonomy supplies the cognitive vocabulary used by StudioCrit \cite{anderson}. Its categories are used as an organising scheme for critique text, not a direct measure of a learner's ability. Six Thinking Hats supplies the inspiration for Critsly's alternative critique perspectives \cite{debono}. Educational LLM research motivates attention to both opportunities and limitations in using generated content for learning \cite{kasneci}. These foundations guide the interface and evidence design without establishing the validity of the particular implementation.

\section{System Design and Implementation}
\subsection{From artefact context to critique and action}
The system follows a layered workflow. A visual board captures design artefacts and intentions. A board-state interface serialises relevant context, annotations and critique history. AI orchestration then routes requests through bounded activities such as reflection, evidence seeking, perspective-taking, synthesis and action planning. The workspace presents the response alongside the artefact, while an evidence layer retains traces for later review.

\begin{figure}[htbp]
\centering\includegraphics[width=\linewidth]{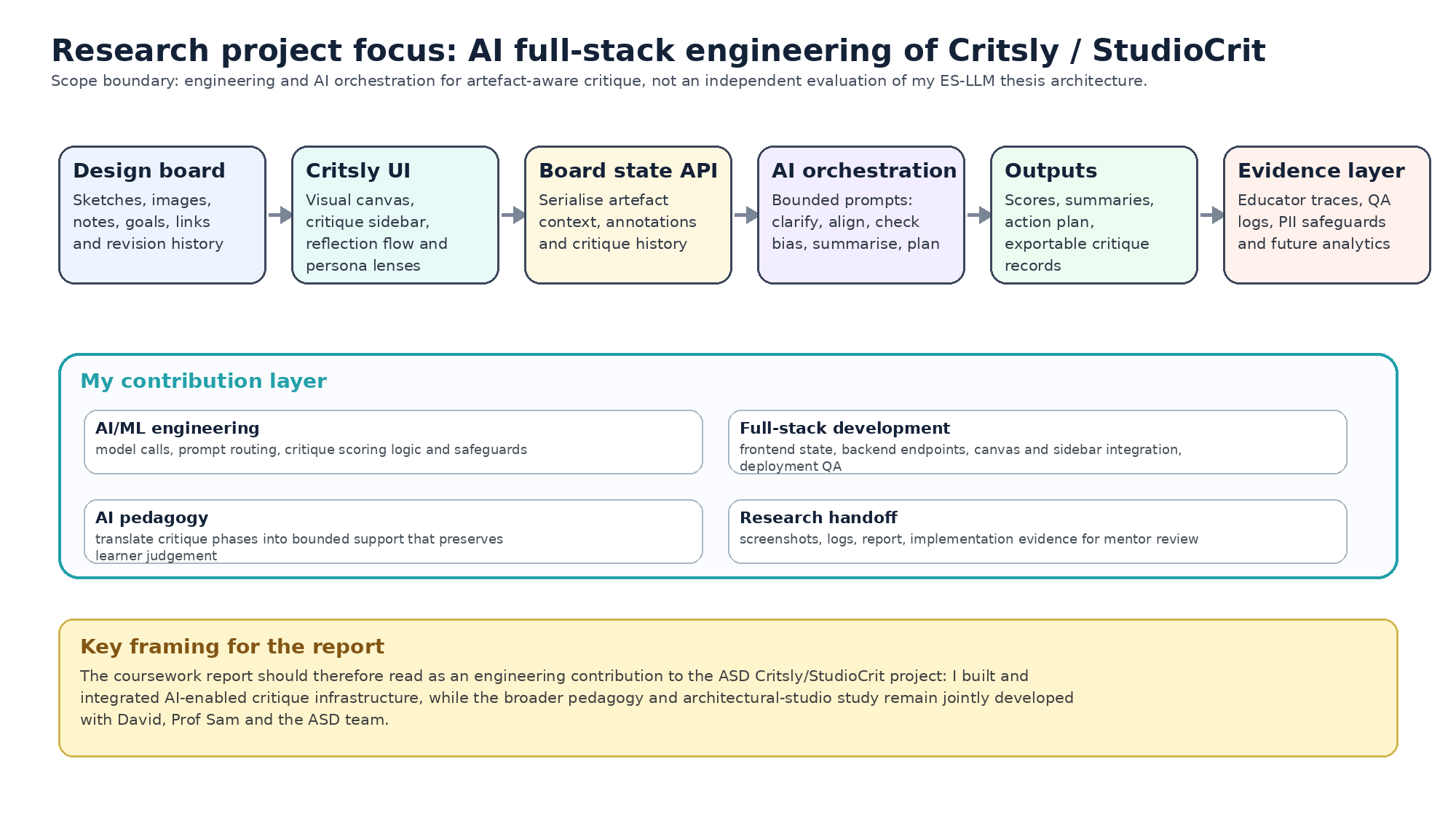}
\caption{Critsly/StudioCrit layered system architecture and the author's engineering contribution boundary. The original diagram distinguishes engineering integration from the broader collaborative pedagogy and architecture-studio study.}\label{fig:architecture}
\end{figure}

Figure~\ref{fig:architecture} reproduces the original architecture and contribution-boundary diagram. All figures in this paper are the original embedded images from the July project report \cite{report}, retained as historical implementation evidence.

Critsly includes projects and boards, design-intention fields, sticky notes, media, feedback pins, an AI critique panel and export controls. Reflecture provides a guided reflection flow. Persona lenses offer different perspectives, and board synthesis can lead to an action plan. The intended pedagogical sequence is to state an intention, inspect the work, consider critique and decide what to revise. Generating an action plan is not evidence that a learner enacted it or benefited from it.

Artefact awareness here means access to structured board context. The report does not establish complete visual understanding of every image or drawing, nor does it measure grounding accuracy. Likewise, the presence of prior critique in the context does not independently establish consistency or correctness across responses.

\subsection{StudioCrit research mode}
StudioCrit is an optional board/class mode. It classifies sticky notes and critique comments using Bloom's revised taxonomy and architectural evaluation dimensions. The reported cognitive levels are Remembering, Understanding, Applying, Analysing, Evaluating and Creating. Architectural categories include Communication, Research, Design and Technical, with more specific subcriteria.

Visible badges and analytics expose assigned labels together with metadata such as matched verbs or terms, confidence and ambiguity. These metadata make the classification inspectable, but they are not evidence of calibrated probabilities or validated explanations. Multi-verb and multi-topic comments can create ambiguity. The submitted report does not provide a complete classifier specification, exact model/configuration, thresholds, taxonomy contents or a gold-standard annotation set; reproducibility and validity are correspondingly limited.

Studios/classes can link multiple boards, supporting shared pin-up, individual/team review and revision/action-plan settings. Educators manage classes and invitations; learners join and contribute to boards; researcher access supports class analytics and export within role boundaries. Table~\ref{tab:scope} summarises the implemented scope documented in the original report.

\begin{figure}[htbp]
\centering\includegraphics[width=\linewidth]{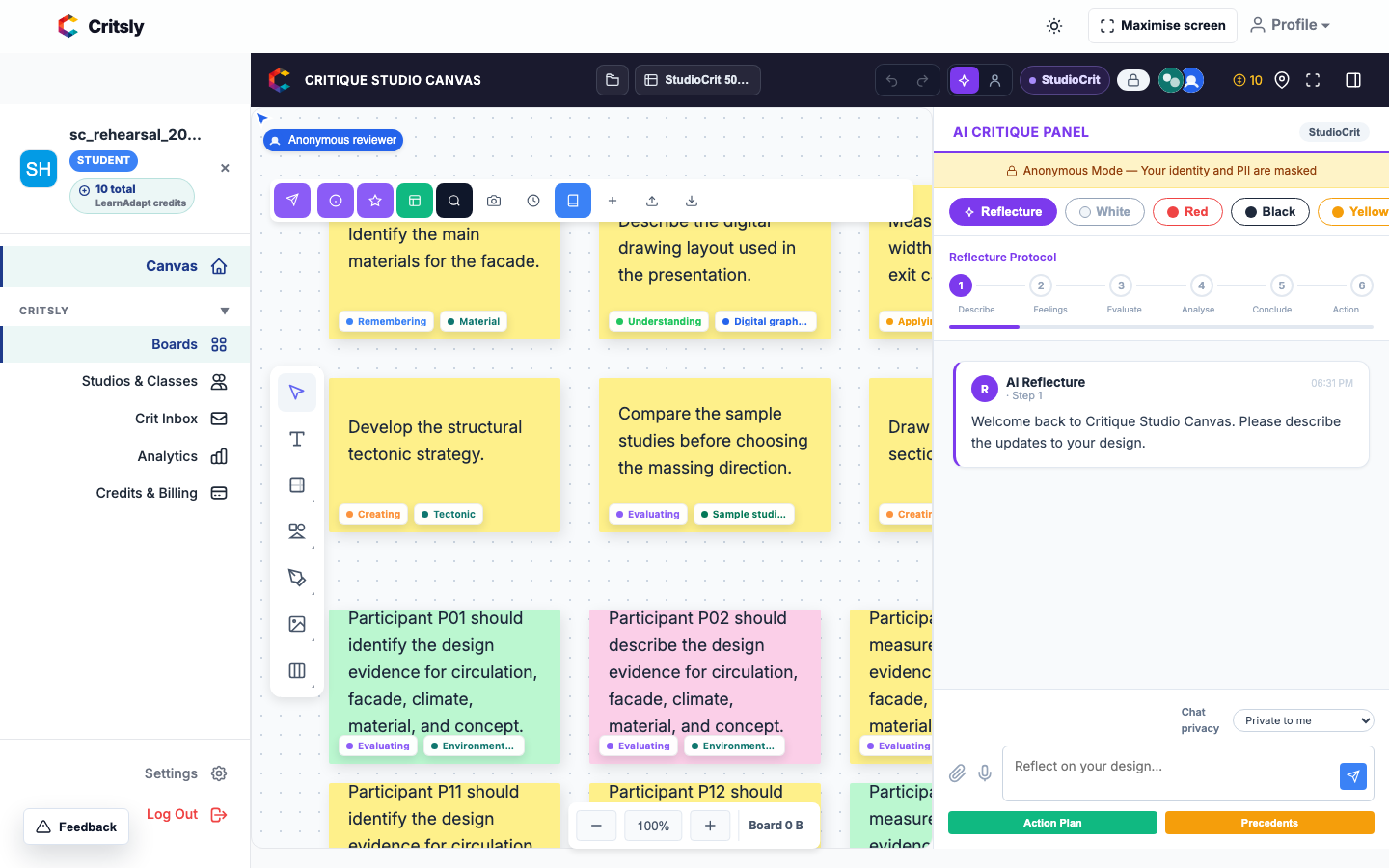}
\caption{StudioCrit active board with classified sticky notes and the AI critique panel. The screenshot shows a disposable rehearsal account and simulated notes.}\label{fig:board}
\end{figure}

\begin{figure}[htbp]
\centering\includegraphics[width=\linewidth]{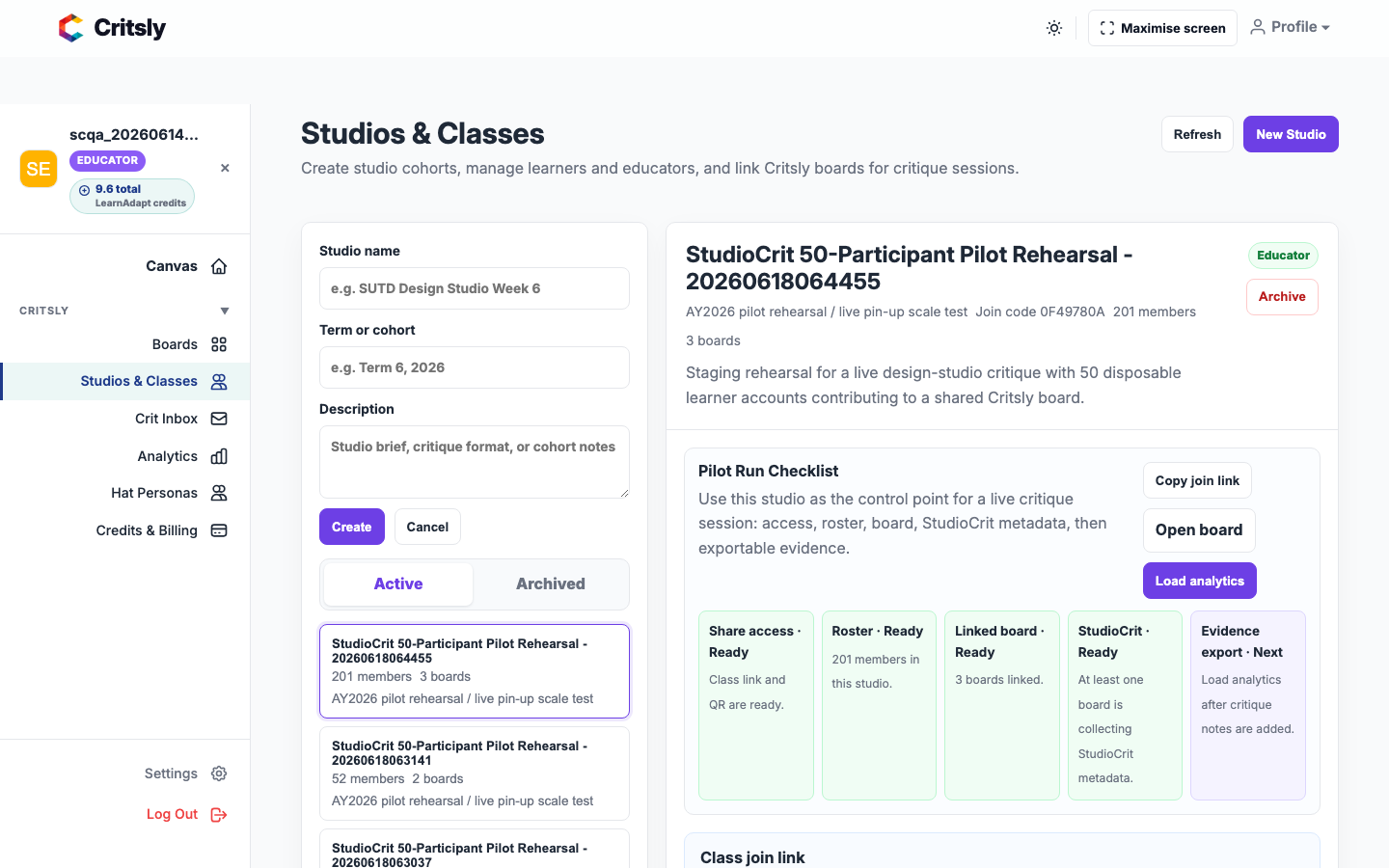}
\caption{Educator StudioCrit class-management view, including invitation controls, linked boards and readiness checks. Historical roster counts shown in the interface should not be interpreted as the sample size of the 50-account rehearsal.}\label{fig:class}
\end{figure}

Figures~\ref{fig:board} and~\ref{fig:class} show the active board and class-management interface. Historical interface labels and roster totals are preserved in these screenshots; the roster totals are not the number of analysed evidence rows or recruited research participants.

\begin{table}[htbp]
\centering\small
\caption{Implemented scope documented in the July 2026 source report.}\label{tab:scope}
\begin{tabularx}{\linewidth}{@{}p{.22\linewidth}Y@{}}
\toprule Layer & Reported functionality \\ \midrule
Workspace & Visual canvas, board toolbar, AI panel, StudioCrit toggle, class management and exports.\\
Critique scaffolding & Design-intention anchoring, Reflecture, persona lenses and critique-to-action planning.\\
Classification & Bloom and architectural labels, matched terms, confidence, ambiguity and taxonomy-version metadata.\\
Research workflow & Studios/classes, linked boards, role permissions, evidence snapshots, CSV export manifest and coder-review fields.\\
Deployment hardening & Polling jitter, failure-aware backoff, hidden-tab throttling, guarded presence updates and production smoke checks.\\
\bottomrule
\end{tabularx}
\end{table}

\subsection{Evidence export and human review}
The refreshed export schema is reported to contain 39 columns spanning studio/board context, provenance, note content and position, machine classification and human review. Examples are shown in Table~\ref{tab:export}. The table is a grouped selection of fields rather than a reconstructed full schema.

The intended research workflow is to preserve a machine-evidence snapshot, obtain independent human codes, reconcile disagreements and export the reviewed record. Manual review should be stored alongside the original label so that a later analyst can distinguish machine output from the agreed human code. The available report documents this workflow and its fields, but does not report a completed dual-coder study.

\begin{table}[htbp]
\centering\small
\caption{Selected export fields reported in the 39-column schema.}\label{tab:export}
\begin{tabularx}{\linewidth}{@{}p{.22\linewidth}Y@{}}
\toprule Field group & Examples \\ \midrule
Studio/board & \texttt{studio\_id}, \texttt{studio\_title}, \texttt{board\_id}, \texttt{board\_title}, \texttt{board\_role}, \texttt{board\_studiocrit\_enabled}.\\
Provenance & \texttt{created\_by\_user\_id}, \texttt{updated\_by\_user\_id}, \texttt{research\_participant\_code}, \texttt{source\_app}, \texttt{artifact\_origin}, \texttt{activity\_frame}.\\
Classification & \texttt{primary\_brt\_level}, \texttt{brt\_levels}, \texttt{architectural\_category}, \texttt{subcriteria}, \texttt{confidence}, \texttt{ambiguity\_flag}, \texttt{taxonomy\_version}.\\
Human review & \texttt{manual\_override\_status}, \texttt{coder\_reconciliation\_status}, \texttt{coder\_notes} and override fields.\\
\bottomrule
\end{tabularx}
\end{table}

\section{Evaluation Method and Evidence Provenance}
\subsection{Simulation scenarios}
The evaluation was a simulation-and-readiness exercise. Three higher-education design-studio scenarios were modelled: a foundation sustainable pavilion studio, an urban mobility spine studio and a facade climate/tectonics lab. Each used multiple linked boards to represent different critique and revision settings. The reported analytical unit was a classified evidence row, not a distinct human participant.

A separate staging rehearsal used 50 disposable learner accounts to exercise a larger studio pin-up workflow. It covered joining, board access, role permissions, note/evidence generation, analytics refresh and export. Throughout this paper, ``50-account rehearsal'' refers to this synthetic account-based exercise; it must not be read as an empirical study of 50 recruited learners.

\subsection{Evidence and analysis boundaries}
The original report drew on CSV exports, class analytics, interface screenshots, role-permission checks and production-hardening records. It reported row counts, cognitive distributions, architectural focus, export fields and operational observations. Ambiguity and low-confidence rows were identified as relevant measures, but the submitted report did not provide numerical rates for them.

This adaptation checks arithmetic and consistency against the text and figures of the submitted PDF. It does not claim access to a complete frozen row-level dataset, source-code revision for every experiment, exact load-generation script or raw timing series. Consequently, summary counts are documentary evidence, not independently rerun measurements. Account counts, evidence rows and completed sessions are different units and are not pooled.

Higher-order categories comprise Analysing, Evaluating and Creating; lower-order categories comprise Remembering, Understanding and Applying. Their proportions describe assigned labels in the simulated corpus. No hypothesis test, treatment comparison, classifier accuracy estimate or causal educational effect is inferred from these proportions.

\section{Findings}
\subsection{Workflow operation}
The source report records successful studio/class organisation with multiple boards, visible cognitive/architectural badges and role-aware access. Students were denied analytics access during the rehearsal checks. Evidence export was assessed as pass/partial because the classification and provenance fields were available while final research coding still required human review. Realtime scale was assessed as partial: the reported account rehearsal passed, but broader public use warranted further event-driven realtime development.

These checks show that exercised paths operated in the reported environment. They do not constitute an exhaustive authorisation audit, a usability evaluation or a service-level guarantee. In particular, successful denial of one protected workflow cannot establish the security of every endpoint.

\subsection{Cognitive-label distributions}
The three-studio simulation produced 109 evidence rows, of which 85 were assigned to higher-order categories (78.0\%) and 24 to lower-order categories (22.0\%). The 50-account rehearsal produced 56 rows, with 46 higher-order assignments (82.1\%) and 10 lower-order assignments (17.9\%). Percentages are recalculated from the original counts. Table~\ref{tab:bloom} transcribes the category counts shown in the source figures and verifies that they sum to the stated totals.

\begin{figure}[htbp]
\centering
\begin{minipage}[t]{.49\linewidth}\centering
\includegraphics[width=\linewidth]{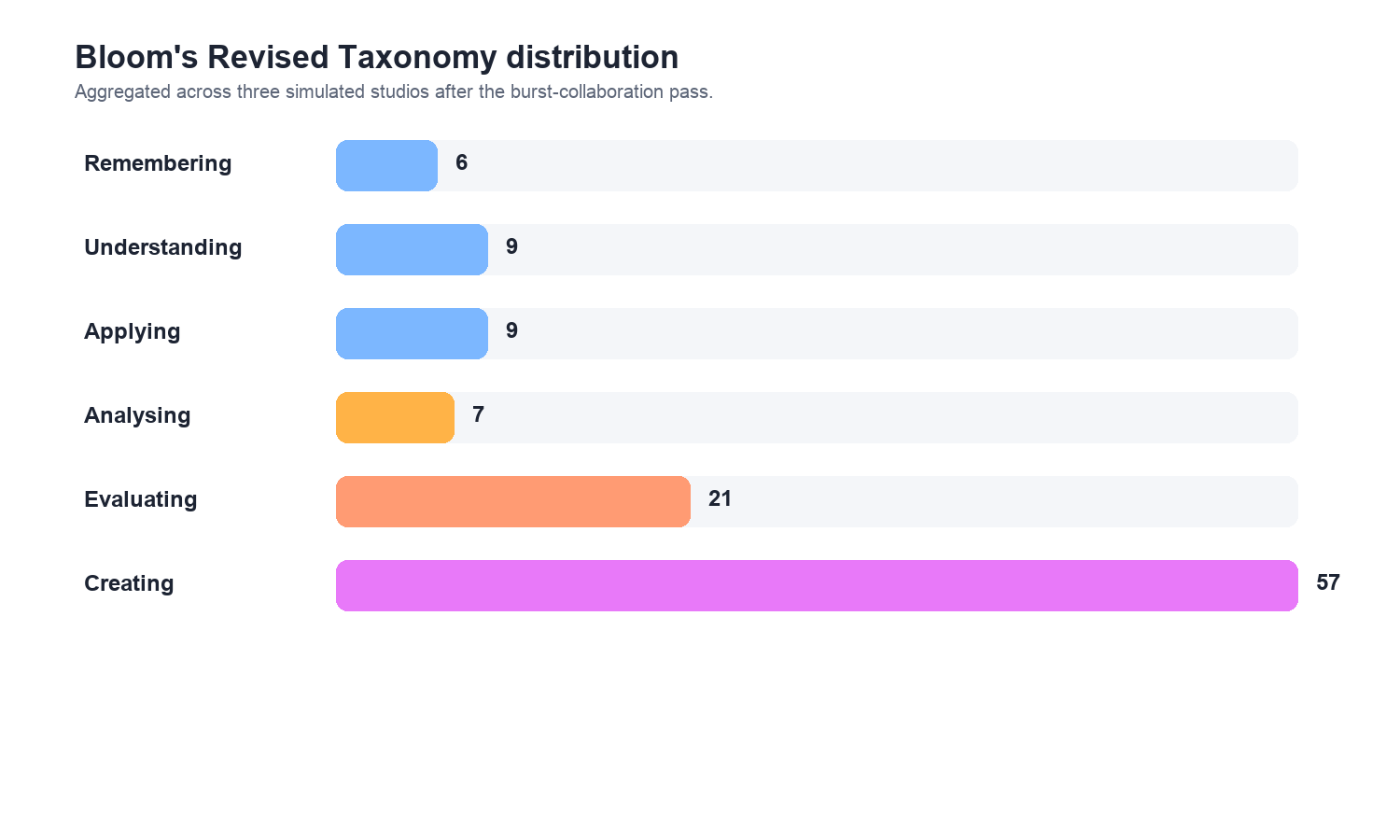}
\small (a) Three simulated studios.
\end{minipage}\hfill
\begin{minipage}[t]{.49\linewidth}\centering
\includegraphics[width=\linewidth]{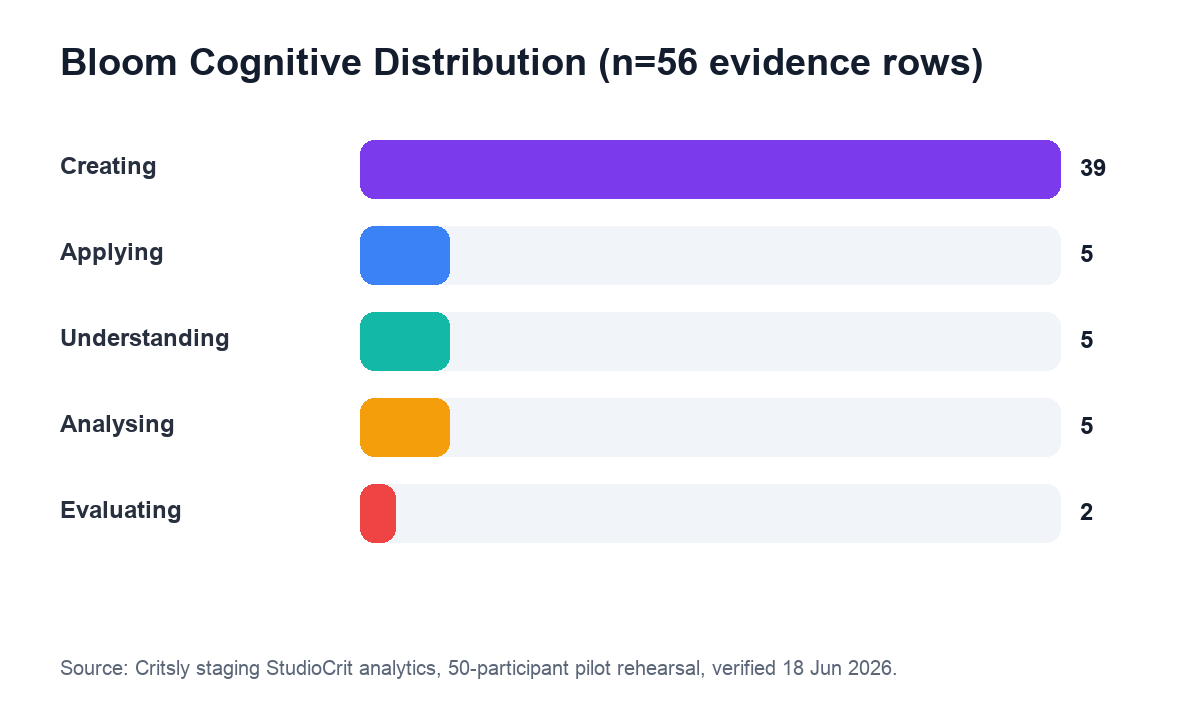}
\small (b) 50-account rehearsal.
\end{minipage}
\caption{Original Bloom-category distributions. Category order differs between the source charts. Counts describe machine-labelled simulated evidence, not measured human cognitive performance.}\label{fig:bloom}
\end{figure}

\begin{table}[htbp]
\centering
\caption{Automated Bloom-label counts from simulated evidence. Counts are rows, not people.}\label{tab:bloom}
\begin{tabular}{@{}lrr@{}}
\toprule Category & Three simulated studios & 50-account rehearsal \\ \midrule
Remembering & 6 & 0\\
Understanding & 9 & 5\\
Applying & 9 & 5\\
Analysing & 7 & 5\\
Evaluating & 21 & 2\\
Creating & 57 & 39\\ \midrule
Total & 109 & 56\\
Higher-order subtotal & 85 & 46\\
Lower-order subtotal & 24 & 10\\
\bottomrule
\end{tabular}
\end{table}

The Creating category was the most frequent in both reported distributions. This is consistent with the types of prompts and notes in the simulated design tasks, but their construction may itself explain the distribution. Without human coding, the figures cannot determine how well the labels represent a note's meaning, much less the cognitive activity or learning of a person.

\subsection{Architectural focus and coverage}
In the 50-account rehearsal, the displayed categories were Technical (24), Design (17), Research (10) and Communication (5), summing to 56. The source interprets the prominence of Technical notes in light of the simulated board's focus on facades, climate, details and materials.

The three-studio architectural chart displays Technical (44), Design (29), Research (21) and Communication (10), which sum to 104. This differs from the 109-row cognitive total. The submitted report does not explain whether the discrepancy reflects missing architectural codes, filtering or another coverage rule. These counts are therefore retained only as the chart's displayed category counts; no five-row category or missing-data mechanism is inferred. Percentages using 109 as their denominator would require reconciliation against the original export. This coverage limitation does not alter the internally consistent Bloom totals, but it prevents a complete joint-distribution analysis.

\begin{figure}[htbp]
\centering
\begin{minipage}[t]{.49\linewidth}\centering
\includegraphics[width=\linewidth]{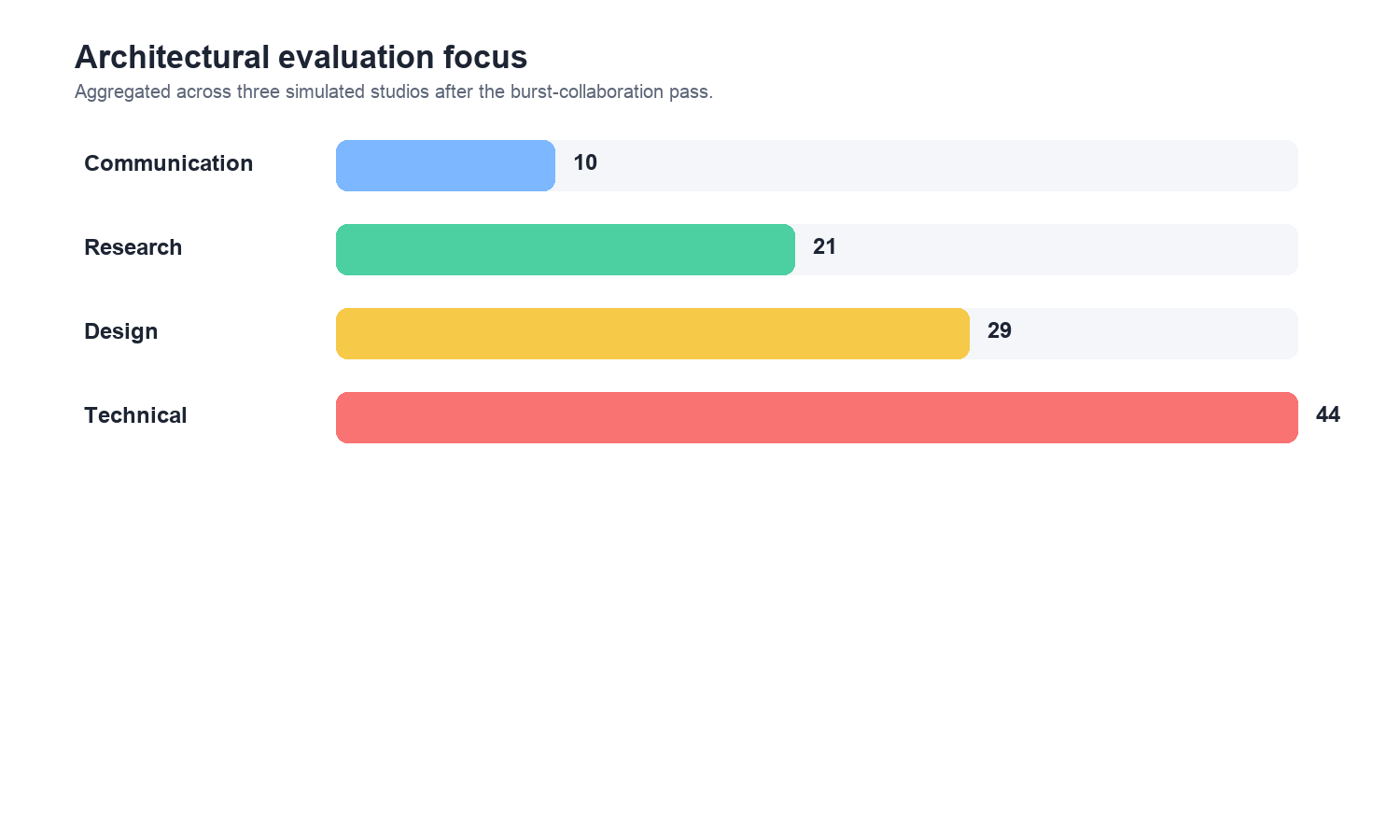}
\small (a) Three simulated studios.
\end{minipage}\hfill
\begin{minipage}[t]{.49\linewidth}\centering
\includegraphics[width=\linewidth]{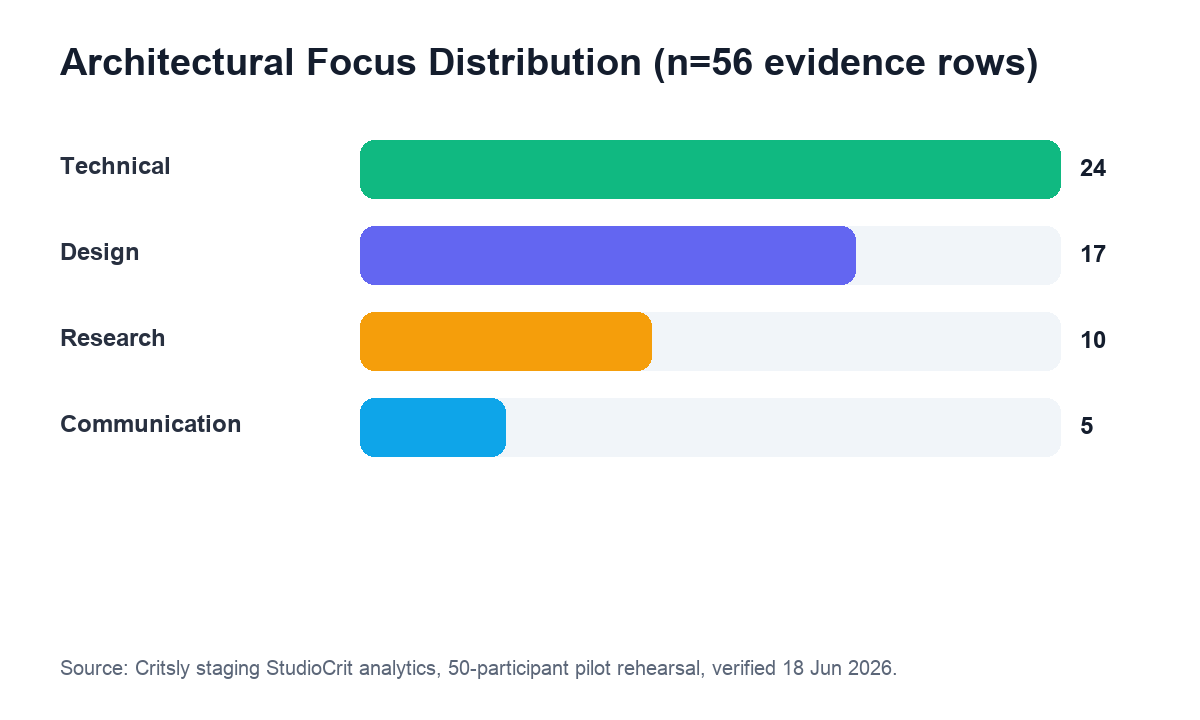}
\small (b) 50-account rehearsal.
\end{minipage}
\caption{Original architectural-focus charts. Displayed counts sum to 104 in (a) and 56 in (b); the unexplained five-row coverage difference in (a) is discussed in the text.}\label{fig:architectural-focus}
\end{figure}

The dashboard also provides word clouds and a cognitive--architectural matrix. Such views can help educators locate concentrations of critique topics, but the visual prominence of a category is not an independently validated measure of critique quality.

\begin{figure}[htbp]
\centering\includegraphics[width=\linewidth]{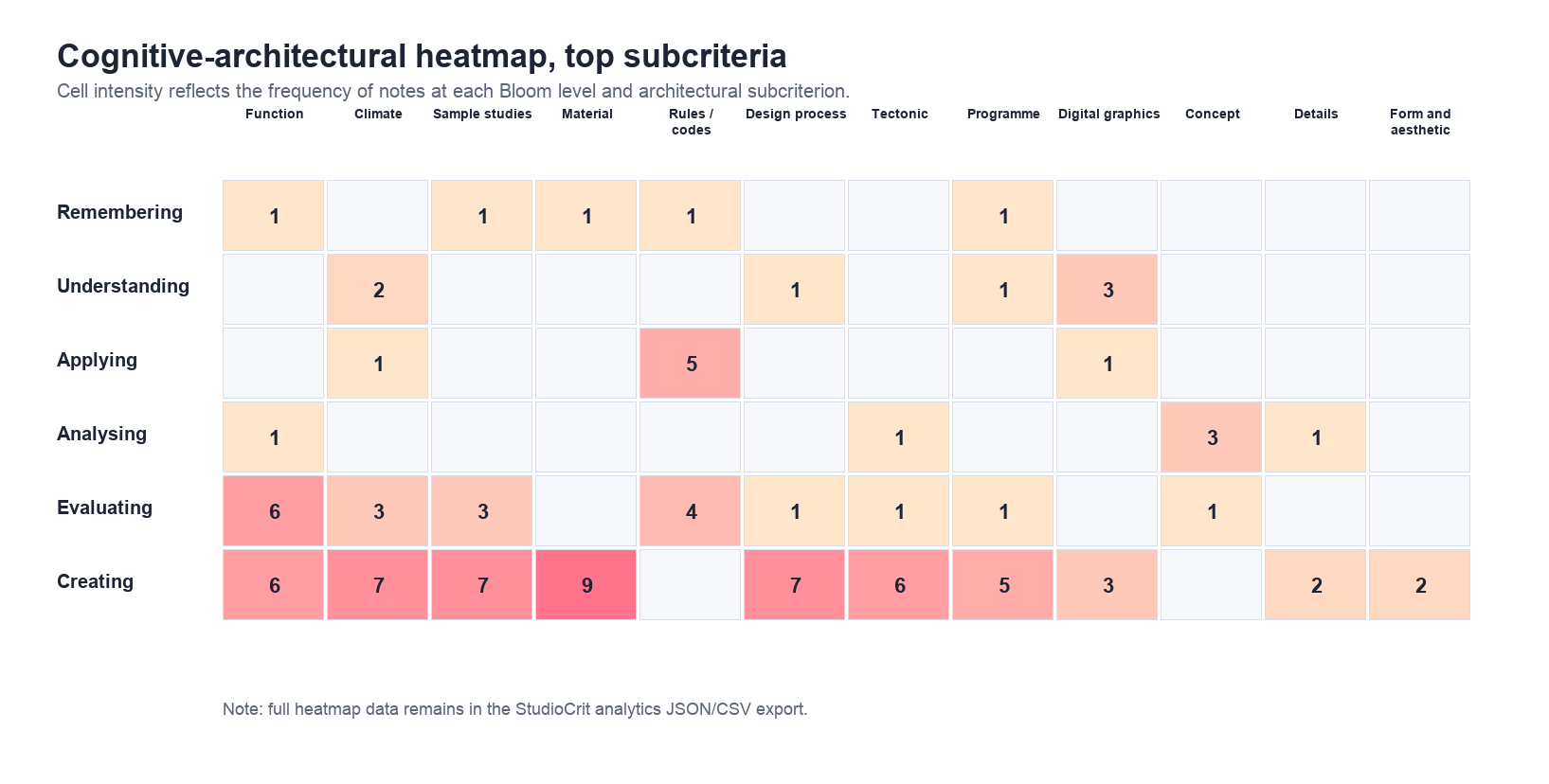}
\caption{Original cognitive--architectural heatmap of the displayed subcriteria. Cell counts are reproduced as shown; the complete joint dataset and exclusion rules were not recovered for this adaptation.}\label{fig:heatmap}
\end{figure}

\begin{figure}[htbp]
\centering\includegraphics[width=\linewidth]{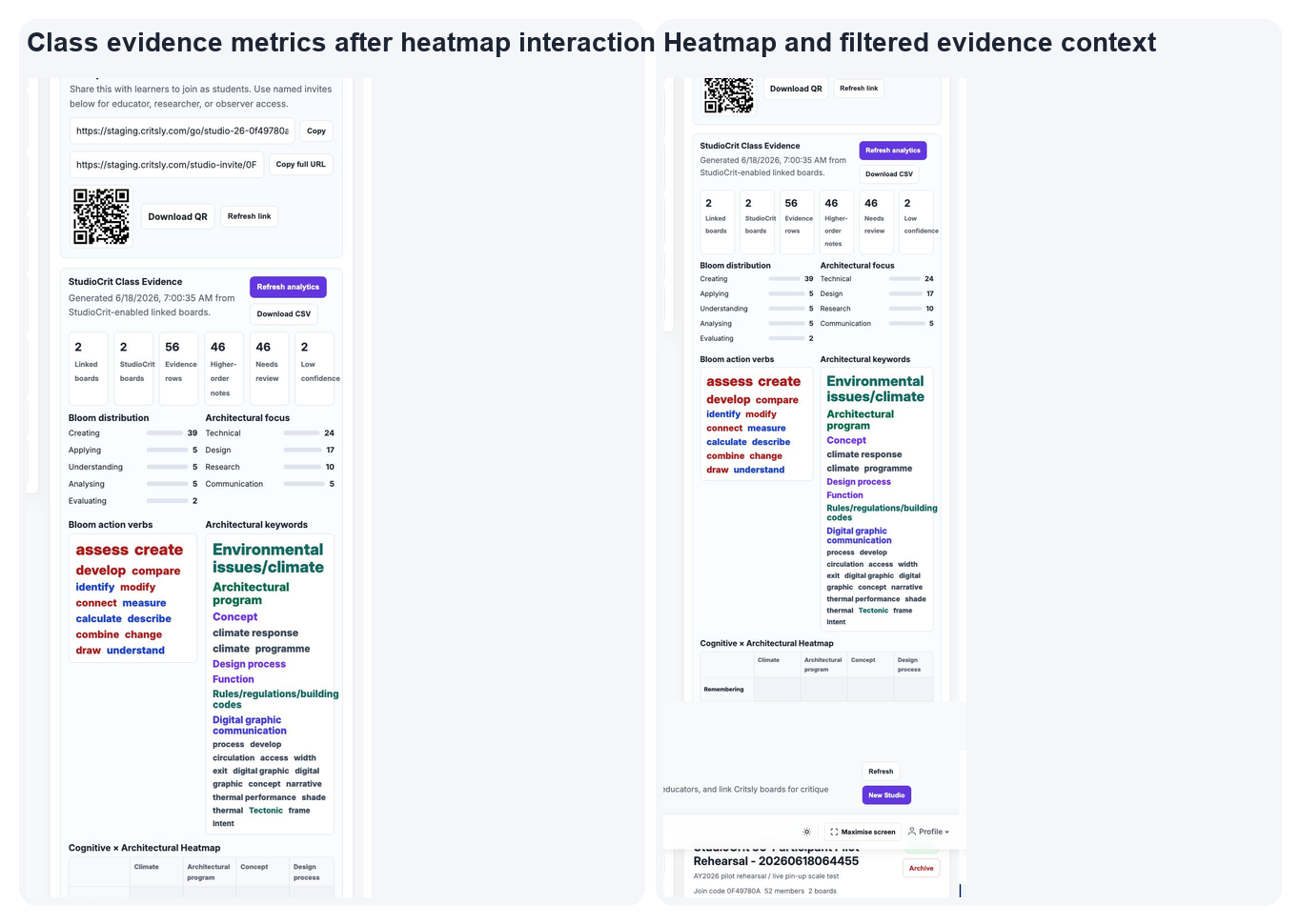}
\caption{Original class-analytics view with word clouds, heatmap context, evidence counts and review indicators. This historical staging screenshot retains its original interface wording and invitation controls.}\label{fig:analytics}
\end{figure}

Figures~\ref{fig:heatmap} and~\ref{fig:analytics} retain the original heatmap and dashboard evidence. The rehearsal dashboard displays 46 rows marked ``Needs review'' and two marked ``Low confidence'' among 56 evidence rows. These UI counts are not estimates of classification error; their definitions, possible overlap and row-level derivation require the underlying export.

\subsection{Deployment hardening}
A later hardening pass added per-tab polling jitter, backoff after failures, hidden-tab throttling and guarded presence updates. The source records verification on staging with 50 distinct learner accounts, followed by promotion to the Critsly production service. Table~\ref{tab:hardening} reproduces the reported operational observations.

\begin{table}[htbp]
\centering
\caption{Hardening-rehearsal observations recorded in the source report.}\label{tab:hardening}
\begin{tabular}{@{}lr@{}}
\toprule Observation & Reported value \\ \midrule
Distinct learner accounts & 50\\
Completed sessions & 50\\
Successful board pulls & 50\\
Student analytics-access denials & 50\\
API session median (p50) & 3.064 s\\
API session maximum & 6.023 s\\
\bottomrule
\end{tabular}
\end{table}

The report identifies production revision \texttt{critsly-php-production-00148-6f4} and states that no severity-at-least-ERROR log entries were reported in the deployment window. The window duration, precise timing endpoints, concurrency schedule, device/network conditions and raw session-level observations are not specified in the submitted report. The timings therefore describe the recorded rehearsal statistic, not per-request latency or sustained classroom capacity. The absence of reported error logs in that window does not establish error-free operation outside it.

\begin{figure}[htbp]
\centering\includegraphics[width=\linewidth]{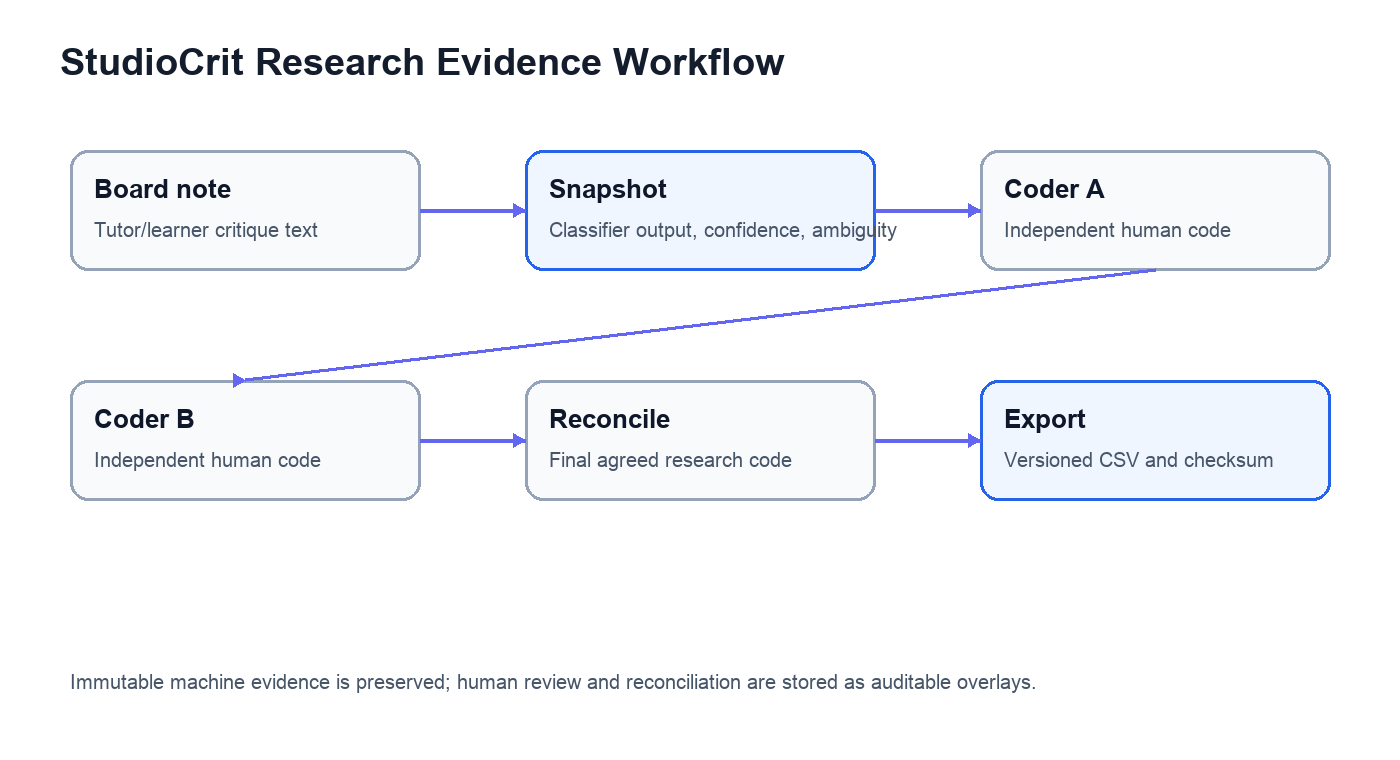}
\caption{StudioCrit research-evidence workflow from the original report: snapshot, independent Coder A/Coder B review, reconciliation and export. The diagram represents the intended review process; no completed human-coder validation is claimed.}\label{fig:review}
\end{figure}

\section{Discussion and Limitations}
\subsection{An inspectable critique-to-evidence workflow}
The system contribution is the continuity between an artefact, its critique, an action plan and an inspectable record. Context and traces can make it easier to revisit what was discussed and identify where further teaching support might be useful. StudioCrit adds a way to organise those traces by cognitive and architectural vocabulary while retaining an educator's interpretive role. These are design affordances demonstrated by implementation; their educational benefits remain hypotheses for evaluation.

Readiness must be assessed along several dimensions. The documented workflows and account rehearsal provide evidence for proceeding to a carefully scoped pilot after appropriate checks. They do not establish classifier validity, authentic classroom usability, stable performance under varied network conditions or adequate governance for an unrestricted deployment. Operational success and educational effectiveness require different evidence.

\subsection{Classification validity}
No completed human-coder validation or inter-rater reliability analysis is reported. A future study should specify a coding manual, define the unit of analysis and have independent coders label an appropriate sample before reconciliation. Agreement should be reported before disagreements are resolved. Cohen's kappa provides one established option for nominal labels \cite{cohen}; ordinal and multi-label decisions require measures suited to their respective coding schemes. Any evaluation must distinguish agreement between human coders from agreement of the machine with a reference annotation.

Ambiguous and low-confidence cases merit explicit analysis. Matched verbs are inspectable cues, but their presence need not establish a note's cognitive level. Confidence scores should not be presented as probabilities without calibration evidence. None of the proposed validation activities or reliability statistics is a completed result of this report.

\subsection{Simulation and reproducibility}
Synthetic notes and disposable accounts cannot reproduce the full range of authentic critique, learner behaviour, device conditions or classroom Wi-Fi. The simulations do not support generalisation to learner populations. Neither the relative cognitive percentages nor the session timings provide a basis for comparing educational outcomes between settings.

The account of implementation is limited by the retained documentary evidence. The exact simulation inputs, complete classifier configuration, frozen 39-column export, timing definitions and load-generation protocol would be needed for reproducible reruns. The unexplained architectural coverage difference further limits joint analysis. Future reporting should include these artefacts and a clear record of any exclusions or unmapped labels, subject to appropriate permissions.

\subsection{Privacy and research governance}
Research exports can contain internal user identifiers, note contents and provenance. Participant codes alone do not make such records anonymous when identity mappings or recognisable artefacts remain available. The source recommends separate identity mapping and a retention/deletion procedure. Before any human-participant deployment, the relevant ethics, consent, access and data-handling arrangements must be established. This paper reports simulations and makes no claim that a completed human-subject study received approval or produced these counts. The reproduced figures show historical staging interfaces and synthetic rehearsal material; raw participant-level exports are not deposited with this paper.

\section{Conclusion}
Critsly and StudioCrit combine an artefact-aware AI critique workspace with studio organisation, provisional cognitive and architectural classification, analytics and exportable evidence. The July 2026 research report documents an implemented system and exercised workflows across simulated studios and account-based rehearsals. Its strongest supported contribution is an engineering account of how critique can remain connected to artefacts and produce inspectable research traces. Further controlled evaluation requires human-coder validation, reconciliation of export coverage and reproducible performance testing. The present evidence supports those next steps without establishing learning gains or validated automated assessment.

\section*{Acknowledgements and contribution scope}
This report derives from Nizam Kadir's SMT 99.580 research project submitted to A/Prof Sam Conrad Joyce. The original report lists Prof Ricky Ang and Dr Sumbul Khan as the author's supervisors. The author acknowledges the broader Critsly/StudioCrit collaboration with Juan David Salazar Rodriguez, Sam Conrad Joyce and the architecture/design team, and the coauthored Critsly demo with Salazar Rodriguez and Sumaiyya Ali \cite{critsly}. The engineering contribution described here concerns AI/ML integration and orchestration, full-stack implementation, bounded pedagogical scaffolding and research handoff. Acknowledgement does not imply approval of this adaptation by the named individuals.

Critsly was developed with support from the AIEDx grant awarded by IAIED (UEN 53487769M).

\section*{Data and software availability}
The system is described at \url{https://critsly.com}. The related demonstration paper is publicly available at \url{https://arxiv.org/abs/2607.09673}. This manuscript reproduces summary observations from the submitted coursework report; it does not deposit raw rehearsal records, a frozen runnable software release or a complete research dataset. Availability of the live service should not be interpreted as reproducibility of its July 2026 state.

\section*{AI assistance}
OpenAI Codex assisted with restructuring the author's submitted report, LaTeX preparation, arithmetic checks and reference lookup for this adaptation. No new experimental observations were generated through that assistance.

\end{document}